\documentclass[a4paper,prb,twocolumn,showpacs]{revtex4-1}
\usepackage[dvips]{graphicx}
\usepackage{amsmath,amssymb}
\usepackage{dcolumn}
\usepackage{bm}
\usepackage{color}

\begin{document}
\bibliographystyle{apsrev}

\title{Inverse simulated annealing: Improvements and application to amorphous InSb}

\author{Jan H. Los}
\affiliation{Institute of Physical Chemistry, Johannes Gutenberg University Mainz, Staudinger Weg 7, D-55128 Mainz, Germany}
\author{Silvia Gabardi}
\author{Marco Bernasconi}
\affiliation{Dipartimento di Scienza dei Materiali, Universit\`a di Milano-Bicocca, Via R. Cozzi 53, I-20125, Milano, Italy}
\author{Thomas D. K\"uhne}
\email{tdkuehne@mail.upb.de}
\affiliation{Department of Chemistry, University of Paderborn, D-33098 Paderborn, Germany}
\affiliation{Institute of Physical Chemistry and Center for Computational Sciences, Johannes Gutenberg University Mainz, Staudinger Weg 7, D-55128 Mainz, Germany}

\date{\today}

\begin{abstract}
An improved inverse simulated annealing method is presented to determine the structure of complex disordered systems from \textit{first principles} in agreement with available experimental data or desired predetermined target properties. The effectiveness of this method is demonstrated by revisiting the structure of amorphous InSb.
The resulting network is mostly tetrahedral and in excellent agreement with available experimental data.
\end{abstract}


\maketitle

\section{Introduction}
Amorphous solids are of interest for a large number of technological applications, ranging from optical lenses and waveguides (oxides), to plastics (organic polymers), solar cells (semiconductors), xerography and non-volatile memory devices (chalcogenides) \cite{Zallen,Elliott}. The determination of the atomistic structure of amorphous and glassy materials is, however, still a major challenge as the lack of long range order prevents a full structural characterization from scattering data \cite{Zachariasen,Elliott,Jansen}. Modeling is therefore particularly useful in resolving the structure of amorphous materials. Reverse Monte Carlo (RMC) \cite{McGreevy1,McGreevy2} 
for instance, is a rather popular technique to generate models that are in very good agreement with experimental scattering and diffraction data. While this allows for an efficient and routine modeling of rather complex disordered structures, the resulting models are not necessarily physical sensible. Furthermore, since no information on the potential energy surface is exploited, a variety of different structural models that are very different from each other, but in similarly good agreement with experiment, can be generated \cite{McGreevy2,Evans, UniqueRMC, SoperRMC1, CRMC, DraboldRMC, HRMC, SoperRMC2}. Finite temperature Molecular Dynamics (MD) or Monte Carlo (MC) simulations offer an alternative route to generate glassy models by quenching from the melt using the simulated annealing (SA) algorithm \cite{SA}. However, due to the large number of degrees of freedom of disordered systems, the annealing has to be conducted as slowly as possible and is therefore computationally exceedingly expensive. This is even more pronounced in conjunction with \textit{ab initio} electronic structure calculations, in spite of recent progress \cite{CP2G, CP2Greview}. As a consequence, the attainable quench-rates are typically several orders of magnitude faster than in experiment. 

Inspired by the inverse design scheme of Franceschetti and Zunger \cite{ID}, we have therefore recently devised a novel method for the generation of amorphous models in agreement with available experimental data \cite{Los1}. This method, which we called Inverse Simulated Annealing (ISA), unifies the RMC and SA techniques to minimize the potential energy, as calculated by Density Functional Theory (DFT) \cite{DFTreview}, while concurrently maximizing the overlap with experiment. Employing an electronic structure method, such as DFT, not only ensures that the atomic configurations are at least metastable by relaxing them into a local-energy minimum, but also facilitates to directly include constraints involving electronic structure properties, such as the band-gap or the dielectric constant, to name a few.

In this paper, we elaborate on the original ISA method \cite{Los1}, including improvements in the minimization algorithm, as well as allowing for volume fluctuations at constant pressure \cite{PR}. The resulting modified ISA approach permits to generate amorphous models at a desired target pressure, which is particularly important when the experimental density of the amorphous is either not known, or expected to differ from the theoretical equilibrium density. In fact, 
it had been shown that amorphous models at the theoretical equilibrium density are generally better reproducing the structure of the real system than models generated at the experimental density \cite{Bouzid}.

The predictive power of the present improved ISA method is demonstrated by revisiting the structure of amorphous InSb (a-InSb), a material of interest for application in infrared photodetectors \cite{Zens} and, at the eutectic composition, as a phase change compound in rewritable digital versatile disks (DVD) \cite{Suzuki}. 
Using the improved ISA method in conjunction with the experimental total pair correlation function (PCF), allows to determine the amorphous structure of a-InSb from \textit{first principles} and to elucidate the local coordination.

The remaining of the paper is organized as follows. In the next section, we present some improvements of the minimization procedure developed for the ISA method and the extension of the method itself to include volume fluctuations. Section III is devoted to its application to a-InSb, followed by a conclusion that is in Section IV.

\section{Inverse Simulated Annealing}

\subsection{Canonical ISA method}
In the original canonical ISA approach \cite{Los1}, the atomic structure is determined by minimizing a function of the form
\begin{eqnarray}
\label{Utilde}
\tilde{U}(\mathbf{R}) &=& U(\mathbf{R})
+ \sum_p w_p \left( \chi_{p}(\mathbf{R}) - \chi^{exp}_{p} \right)^2
\end{eqnarray}
by varying the atomic positions $\mathbf{R} = \{ {\mathbf r_i} \}$. In Eq.~\ref{Utilde}, $U(\mathbf{R})$ is the potential energy, while $\chi_{p}(\mathbf{R})$ is the calculated value of a property $p$ and $\chi^{exp}_p$ the experimental reference data. The latter may include structural properties from scattering data, but also properties related to the electronic structure, such as the band gap. Alternatively, we note that in analogy to the inverse design technique \cite{ID}, $\chi^{exp}_p$ could be replaced by a predetermined desired target property for a certain application. 

In minimizing $\tilde{U}(\mathbf{R})$, we take advantage of the fact that by employing Eq.~\ref{Utilde}, the accessible phase space is substantially reduced and confined to energetically low-lying atomic configurations. In other words, although the dimensionality of the phase space is unchanged, the optimization is guided in a funnel-like fashion towards the minimum of $\tilde{U}(\mathbf{R})$. Nevertheless, devising an efficient minimizer is still one of the main challenges of our approach to facilitate the minimization of $\tilde{U}$ from first-principles. 

The hybrid MC-based SA method introduced in our previous work \cite{Los1}, has the following key properties: i) it is a ``fuzzy'' hybrid MC method with all atom trial moves involving all nuclear forces, which ii) is performed in the microcanonical NVE ensemble with a correspondingly modified acceptance probability. The all atom trial moves are generated by a single energy conserving MD step using a slightly modified velocity-Verlet algorithm, where the time step $dt \in (0,dt_{max})$ is chosen at random, while $dt_{max}$ is adjusted on-the-fly to achieve an acceptance probability of $\sim$50\%. With this, $dt_{max}$ is typically up to an order of magnitude larger than the maximum permissible time step in a conventional MD simulation. It is important to note that the velocities are updated even if the configuration itself is rejected. In this way, the velocities ${\bf v}'_i$ are gradually turned into the direction of the forces upon repeated rejections, which results in an increased acceptance probability for large displacements (i.e. large $dt$) and thus significant improvement in efficiency \cite{Los1}. Furthermore, due to the global nature of the stochastic optimization method, trapping in energetically high local minima is avoided. For the purpose to facilitate MC simulations within the NVE ensemble, the acceptance probability of the trial move is given by \cite{Ray}:
\begin{eqnarray}
\label{probNVE1}
P = \min{ \left( 1, \left( \frac{ E - \tilde{U}' }
{ E - \tilde{U} } \right)^{3N/2 -1} \right) },
\end{eqnarray}
where $E$ is the applied total energy and $N$ is the number of atoms. Note that in a NVE simulation, the total energy $E$ is fixed, while
the average temperature is assigned by $E - \tilde{U} = K = \frac{3}{2} N k_{B} T$, where $K$ is the kinetic energy. For a SA simulation within the NVE ensemble, $E$ is gradually reduced from $E_{max}$ to $E_{min}$. For example, in the case of generating an amorphous model by quenching it from the melt, $E_{max}$ should be taken such that the system is in the liquid phase, whereas $E_{min}$ should be chosen as close as possible, though slightly above the ground state energy $\tilde{U}_0$ of the eventual amorphous phase. 

In the following we present some modifications of our original hybrid MC-based optimization method. 
\begin{figure}
\vspace*{0.0cm}
\hspace*{-0.5cm}
\includegraphics[width=9.4cm,clip]{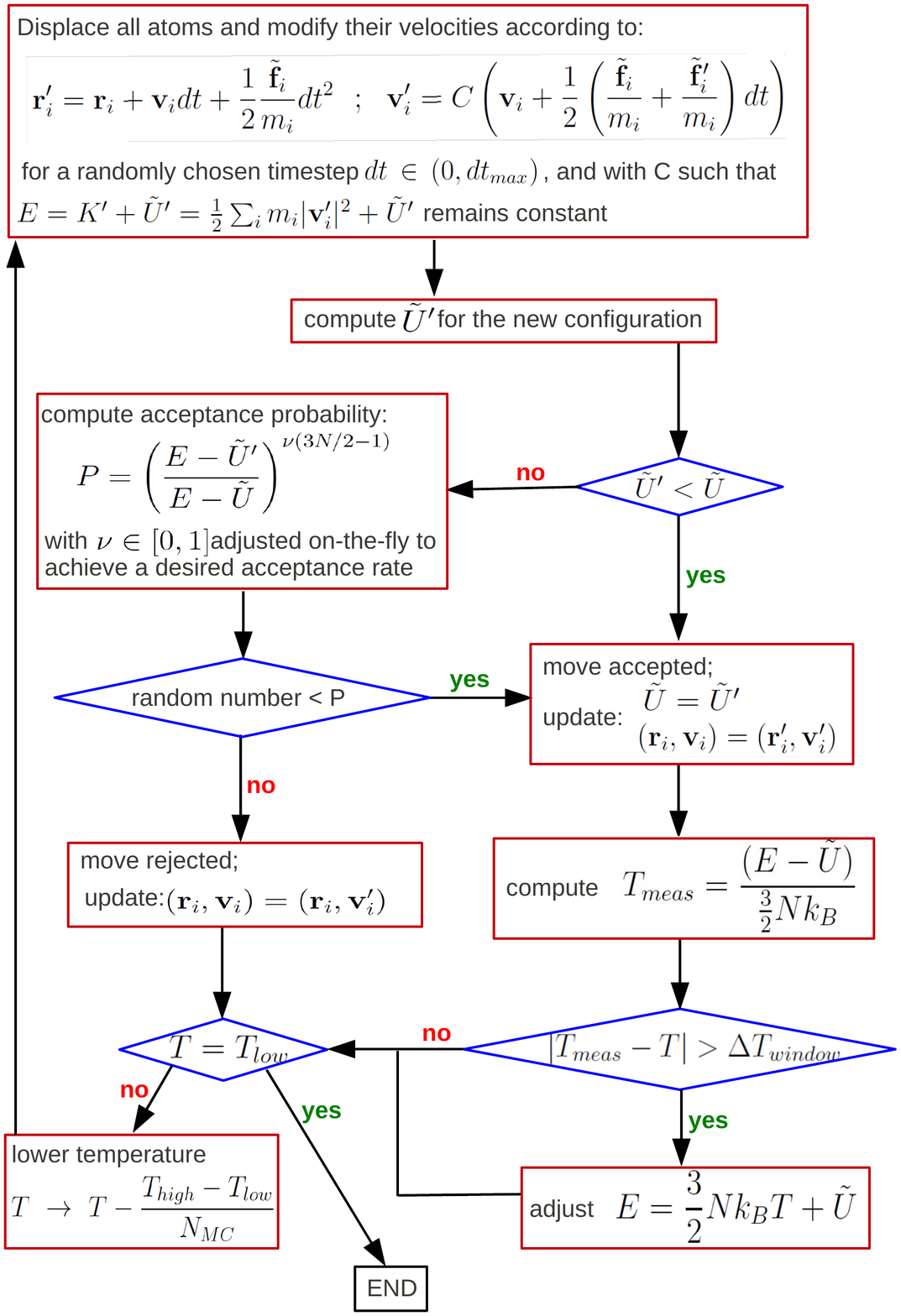}
\label{flowchart}
\caption{\label{flowchart} Flowchart of the minimization algorithm used in conjunction with the ISA method. Therein, $\bf{v}_i$ are the velocities of atom $i$, while $\tilde{\bf f}_i$ corresponds to the total nuclear force $\partial \tilde{U}/\partial {\bf r}_i$.} 
\end{figure}
In this improved version, outlined in the flowchart of Fig.~\ref{flowchart}, the acceptance probability is generalized to:
\begin{eqnarray}
\label{probNVE2}
P = \min{ \left( 1, \left( \frac{ E - \tilde{U}' }
{ E - \tilde{U} } \right)^{ \nu(3N/2 -1)} \right) },
\end{eqnarray}
where $ \nu $ is a number larger than zero and adjusted on-the-fly to achieve the desired acceptance rate of $\sim$50\%, while $dt_{max}$ is now kept constant. In other words, for $\nu \rightarrow 0$ all moves are accepted, while for a very large $\nu$ only downhill moves for which $\tilde{U}' < \tilde{U}$ holds are accepted. 
The value at which $\nu$ equilibrates depends on $dt_{max}$. As it turns out, to maximize the efficiency, $dt_{max}$ should be chosen such that $\nu$ equilibrates to a value around 1/2. This corresponds to a $dt_{max}$, which is about 10 times larger than the typical time step of a MD simulation for the particular system. 

Another improvement regards the annealing schedule. In fact, minimizing Eq.~\ref{Utilde} by reducing $E$ as a function of MC steps from $E_{max}$ to $E_{min} \simeq \tilde{U}_0$ is rather inconvenient, since $\tilde{U}_0$ is \textit{a priori} unknown. 
Therefore, we have chosen to linearly decrease $T$ from $T_{high}$ to $T_{low}$ as a function of MC steps instead of $E$, while maintaining to conduct the minimization in the NVE ensemble as in our original approach. To that extend, $E$ is updated as soon as the instantaneously measured temperature $T_{meas}$ is outside a given window $\Delta T_{window}$ around the applied temperature $T$. The actual temperature $T_{meas}$ is associated to the kinetic energy $K$ of the system by $K = E - \tilde{U} = \frac{3}{2} N k_{B} T_{meas}$. Therefore, after each accepted move, whenever $| T_{meas} - T | > \Delta T_{window}$, $E$ is adjusted such that the kinetic energy, and correspondingly the velocities, are in agreement with the applied temperature, i.e. such that $E - \tilde{U} = \frac{3}{2} N k_{B} T$.

To illustrate the increased efficiency of the modified optimization scheme, we have applied the canonical ISA method to determine low energy structures of amorphous carbon using the reactive LCBOPII carbon potential \cite{LCBOPII}. 
\begin{figure}
\vspace*{0.0cm}
\includegraphics[width=8cm,clip]{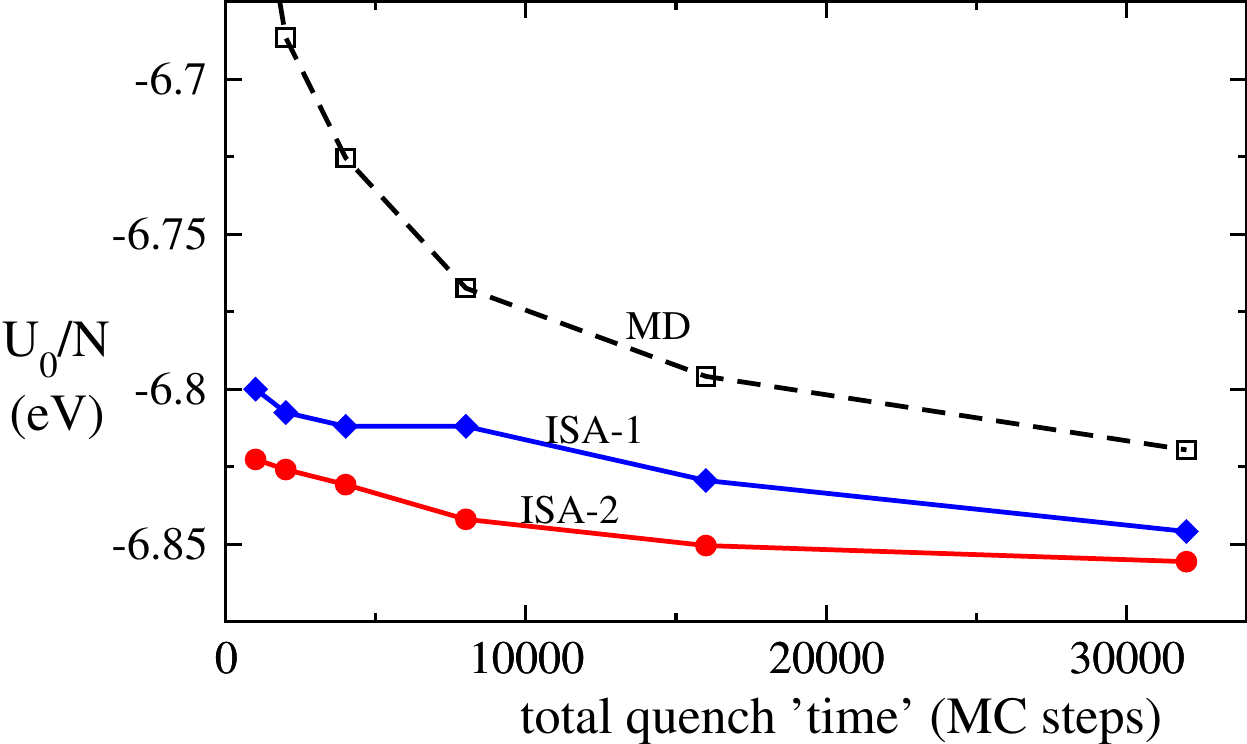}
\caption{\label{u0_nmc} Comparison of the average over the final potential energies per atom $U_0/N$ at 0~K as a function of the total quenching `time'. The amorphous carbon models were generated using the empirical LCBOPII carbon potential \cite{LCBOPII} by a conventional MD-based SA simulation, the original ISA-1 technique as published in Ref.~\onlinecite{Los1}, as well as the improved ISA-2 method, which is outlined in the flowchart of Fig.~\ref{flowchart}. The averages are based on 40 independent simulations.} 
\end{figure}
A comparison of the average final energies per particle $U_0/N$ as a function of the total quenching time' (i.e. number of potential energy evaluations) of the amorphous phases as obtained by the original and modified versions of the ISA technique and those from an usual MD-based SA simulation, is given in Fig.~\ref{u0_nmc}. As can be seen, the improved ISA method constitutes a sizable improvement with respect to the original scheme and is substantially more efficient than the conventional MD-based SA approach. 

\subsection{Isobaric ISA method}

Determining the atomic configuration of a system using the canonical ISA method at constant particle density, will usually lead to structures, whose stress tensor is non-vanishing. In fact, the experimental density is often not known from the outset and it would be in general desirable to generate relaxed amorphous models at the theoretical equilibrium density. The latter might differ from the experimental equilibrium density due to the finite accuracy of the employed level of theory to calculate the interatomic forces. Nevertheless, amorphous models at zero pressure can be directly generated by including volume fluctuations in the ISA scheme, similar to standard MC simulations at constant pressure \cite{BinderMC}. 

In order to keep the pressure fluctuating around a given target pressure $P_a$, we simply add the volume dependent contributions of the Gibbs free energy to Eq.~\ref{Utilde}. The extended objective function $\tilde{G}(\mathbf{R},V)$ now reads as: 
\begin{eqnarray}
\label{Gtilde}
\tilde{G}(\mathbf{R},V) &=& U(\mathbf{R}) - N k_B T ln(V) + P_a V \nonumber \\
&+& \sum_p w_p \left( \chi_{p}(\mathbf{R},V) - \chi^{exp}_{p} \right)^2, 
\end{eqnarray}
where $V$ is the volume of the system, while 
\begin{eqnarray}
\label{probNVE}
P = \min{ \left( 1, \exp{ (-\beta \Delta \tilde{G}) } \right) }
\end{eqnarray}
is the standard MC acceptance probability in the NPT ensemble. The change in $\tilde{G}(\mathbf{R},V)$ due to a combined all atom and volume move is denoted as $\Delta \tilde{G}$, while $\beta=1/(k_B T)$. 
However, minimizing $ \tilde{G}(\mathbf{R},V) $ with respect to $V$ requires to solve:
\begin{eqnarray}
\label{dUtildedV}
\frac{d \tilde{G}}{dV} &=& P_a - P_{vir} - P_{kin} \nonumber \\
&+& 2 \sum_p w_p \left( \chi_{p}(\mathbf{R},V) - \chi^{exp}_{p} \right)
\frac{d \chi_{p} }{dV} \nonumber \\ &=& P_a - P +
2 \sum_p w_p \left( \chi_{p}(\mathbf{R},V) - \chi^{exp}_{p} \right)
\frac{d \chi_{p} }{dV} \nonumber \\ &=& 0, 
\end{eqnarray}
where $P_{vir} = -dU/dV$ is the virial contribution and $P_{kin} = N k_{B} T/V$ the ideal vapor (kinetic) contribution to the total pressure $P = P_{kin} + P_{vir}$. Hence, without the constraint terms, the condition $d\tilde{G}/dV = 0$ implies that $P = P_a$. So in that case we would find the instantaneous pressure $P$ of the system in the NPT ensemble fluctuating around $P_a$. 
But, how to achieve the same at the presence of multiple constraint terms is not obvious since during the simulation $d \chi_{p} /dV$ is in general non-zero, as well as $\chi_{p}(\mathbf{R},V) - \chi^{exp}_{p}$. 
For the purpose to circumvent or at least reduce the spurious pressure contributions that are originating form the various constraint terms, the weight factor $w_p$ should be chosen as small as possible, although at the same time large enough to achieve the desired agreement with the desired target property. 

Fortunately, for certain properties the just mentioned issue can be solved in a more elegant way by defining the constraint terms in a form that is invariant under volume fluctuations. In particular, this is possible for the PCF \cite{PCF}, for which experimental data is very often available. To that extend, the constraint term, which we denote as $\tilde{U}_g$, can be defined in a scale invariant form as:
\begin{eqnarray}
\label{Ut_PCF}
\tilde{U}_g(\mathbf{R}) = w_g
\sum_{n=1}^{N_{r_n}} ( g_s(\mathbf{R}; s_r r_n) - g^{exp} (r_n) )^2, 
\end{eqnarray}
where the sum is over the $N_{r_n}$ grid points $r_n$ on which the experimental PCF $g^{exp} (r_n)$ is discretized. The scale factor $s_r = (V/V_{exp})^{(1/3)} = (\rho_{exp}/\rho)^{(1/3)}$, while $\rho =1/V$ is the actual and $\rho_{exp} = 1/V_{exp}$ the corresponding experimental density of the system. The eventual scaled PCF is denoted as $g_s(\mathbf{R}; s_r r_n)$, where $s_r r_n$ are discretized and rescaled grid points. 
In the case of the reduced PCF (RPCF), which is defined as $G(r) = 4 \pi r \rho (g(r) -1)$, the scale invariant form of the constraint term reads as:
\begin{eqnarray}
\label{Ut_RPCF}
\tilde{U}_G(\mathbf{R}) = w_G
\sum_{n=1}^{N_{rn}} ( s_r^2 G_s(\mathbf{R}; s_r r_n) - G^{exp} (r_n) )^2
\end{eqnarray}
In order to calculate the force contributions from Eqs.~\ref{Ut_PCF} and \ref{Ut_RPCF}, respectively, we have employed a smoothing filter. More details on the Gaussian smoothening and on the choice of the weight factors $w_g$ and $w_G$, respectively, as well as how to eventually calculate the corresponding force contributions are given in the Appendix.

Due to the fact that the scale factor $s_r$ is strictly related to the density ratio $\rho_{exp}/\rho$, for a given $\rho_{exp}$, the scale factor $s_r$ only changes due to volume fluctuations. 
As a consequence, the experimental target density $\rho_{exp}$ must be known. However, in order to deal with cases where $\rho_{exp}$ is \textit{a priori} unknown, we have also implemented a scheme where $s_r$ is varied in addition to the aforementioned volume fluctuations. In principle, this is identical to vary the unknown $\rho_{exp}$ in order to predict the value that maximizes the agreement between the instantaneously computed and desired target PCF. 

\section{Application to amorphous InSb}


In our previous work \cite{Los2}, the model of a-InSb was generated by quenching from the melt using the DFT-based second generation Car-Parrinello MD method of K\"uhne et al. \cite{CP2G, CP2Greview}. It was found that short MD quenches (up to 130~ps) gave rise to an octahedral arrangement, while the resulting structure of a longer MD quench ($\sim$330~ps) was mostly tetrahedral. However, for the latter a much more confined basis set had been used to accommodate for the increased computational cost. Furthermore, in a subsequent study on the closely related In$_3$SbTe$_2$ compound, it was demonstrated that the structure critically depends on the density as well \cite{Los3}. 
While at high density, the bonding of the In atoms is mostly octahedral-like, at low density a sizable fraction of tetrahedral-like geometries had been observed. This variability originates from a close competition between tetrahedral-like and octahedral-like sites, as previously found in related amorphous tellurides \cite{Caravati1, Akola1, Elliott1, CaraPRL, Caravati2, Akola2, DraboldElliott, Caravati3, Matsunaga, Caravati4, Akola3, Kolobov, Mazzarello1, Gabardi, Elliott2, Mazzarello2}. Nevertheless, this immediately suggests that in the presence of nanovoids in the amorphous/liquid phase, the inclusion of van der Waals interactions might be particularly important and may result is a somewhat higher density \cite{Spreafico}. In fact, neglecting van der Waals interactions, at the DFT level using the Perdew-Burke-Ernzerhof (PBE) exchange and correlation (XC) functional \cite{PBE}, the theoretical equilibrium density was shown to be 3 \% lower than the experimental value \cite{Los2, Shevchik}. 

Utilizing the increased efficiency of the modified ISA method, we are revisiting here the amorphous phase of InSb including van der Waals interactions in order to investigate the sensitivity of the structure on the density and in particular if it is more tetrahedral or octahedral.

\subsection{Computational Details} 

To model the amorphous phase of InSb a cubic supercell consisting of 216 atoms subject to periodic boundary conditions was considered. 
In the following two ISA simulations are presented, a canonical calculation at the experimental density and an isobaric one at zero pressure. In both cases the RPCF had been employed as the experimental target function to guide to optimization, while $w_G$ of Eq.~\ref{Ut_RPCF} has been taken equal to 1.0 eV \AA$^{-4}$. The experimental $G(r)$ as well as the corresponding density $\rho_{exp}$ were both obtained from the data in Ref.~\onlinecite{Shevchik} on a-InSb grown by sputtering. 
For the constant volume simulations the parameter $s_r$ was fixed and equal to 1, while at at constant pressure it follows the density fluctuations through $s_r = \rho_{exp}/\rho$. Starting from a well equilibrated melt, the system was quenched to 1000~K and re-equilibrated for around 1000 hybrid MC-steps \cite{HMC1, HMC2}. Thereafter, the temperature was linearly decreased from 1000~K to 300~K within 8000 hybrid MC-steps before cooling the system to the ground state in additional 1000 hybrid MC-steps. 

For the purpose to compute the potential energies and nuclear forces at the semi-local DFT level, we have linked our ISA code with the \textsc{Quickstep} module of the CP2K suite of programs \cite{VandeVondele}. In this method, the Kohn-Sham orbitals are expanded in contracted Gaussians, whereas the electronic charge density is represented using plane waves. For the former, an accurate double-$\zeta$ valence polarized basis set (DZVP) was employed \cite{MoloptBasis}, while the latter was expanded on a regular plane wave grid using a density cutoff of 100~Ry to efficiently solve the periodic Hartree potential. Moreover, the PBE XC functional \cite{PBE} and norm-conserving Goedecker-type pseudopotentials with three and five valence electrons for the In and Sb atoms were used \cite{GTH, HGH, KrackPP}. Due to the presence of disorder, the Brillouin zone integration was restricted to the supercell $\Gamma$-point only. In the constant pressure simulations we also included a damped interatomic potential to approximately account for van der Waals interactions \cite{Grimme}. 

For the sake of comparison we have additionally performed two rather long DFT-based MD quenches (225~ps and 350~ps, which are denoted as MD-1 and MD-2, respectively) with exactly the same settings as before using the second-generation Car-Parrinello approach together with a discretized integration time step of 2.0~fs. 

\subsection{Canonical ISA Simulation}


A comparison of the final energies of the present canonical ISA simulation and DFT-based MD quenches, as well as the previously published model of Ref.~\onlinecite{Los2} are given in Table~\ref{Table1}. 
\begin{table}
\begin{tabular}{cccc}
Model &  Quenching time&  Energy & Pressure \\
    &  (ps/$N_{HMC})$ & (eV) & (GPa) \\ \hline  \hline
MD-1&  $\sim$ 225 ps &   3.715  & -0.67 \\
MD-2&  $\sim$ 350 ps &  -1.548  & -0.87\\
MD \cite{Los2}  &  $\sim$ 330 ps &  1.004 & 0.67\\
ISA & 10000  &  0.000  & 0.30 \\ \hline  \hline
\end{tabular}
\caption{\label{Table1}Potential energies of the geometry optimized models of a-InSb relative to the one of the canonical ISA method, as generated by two long DFT-based MD quenches (MD-1 and MD-2) and the previously published model of Ref.~\onlinecite{Los2} (MD). The corresponding pressures of the optimized models at the experimental density are given in the last column.}
\end{table}
In each case, the potential energies and pressures have been computed at their respective nuclear ground state, i.e. after a geometry optimization. From Table~\ref{Table1} it is apparent that the eventual ground state energies are the lower the longer the quenching time. However, even more interestingly, it also demonstrates that the modified ISA method is energetically very competitive with even the longest MD quenches, despite the presence of constraints that can only increase the potential energy. In addition, for the particular example, the modified ISA technique is computationally at least a factor 15 more efficient than an equivalent DFT-based MD quench. 

\begin{figure}
\vspace*{0.0cm}
\hspace*{-0.4cm}
\label{fig3}
\includegraphics[width=8.cm,clip]{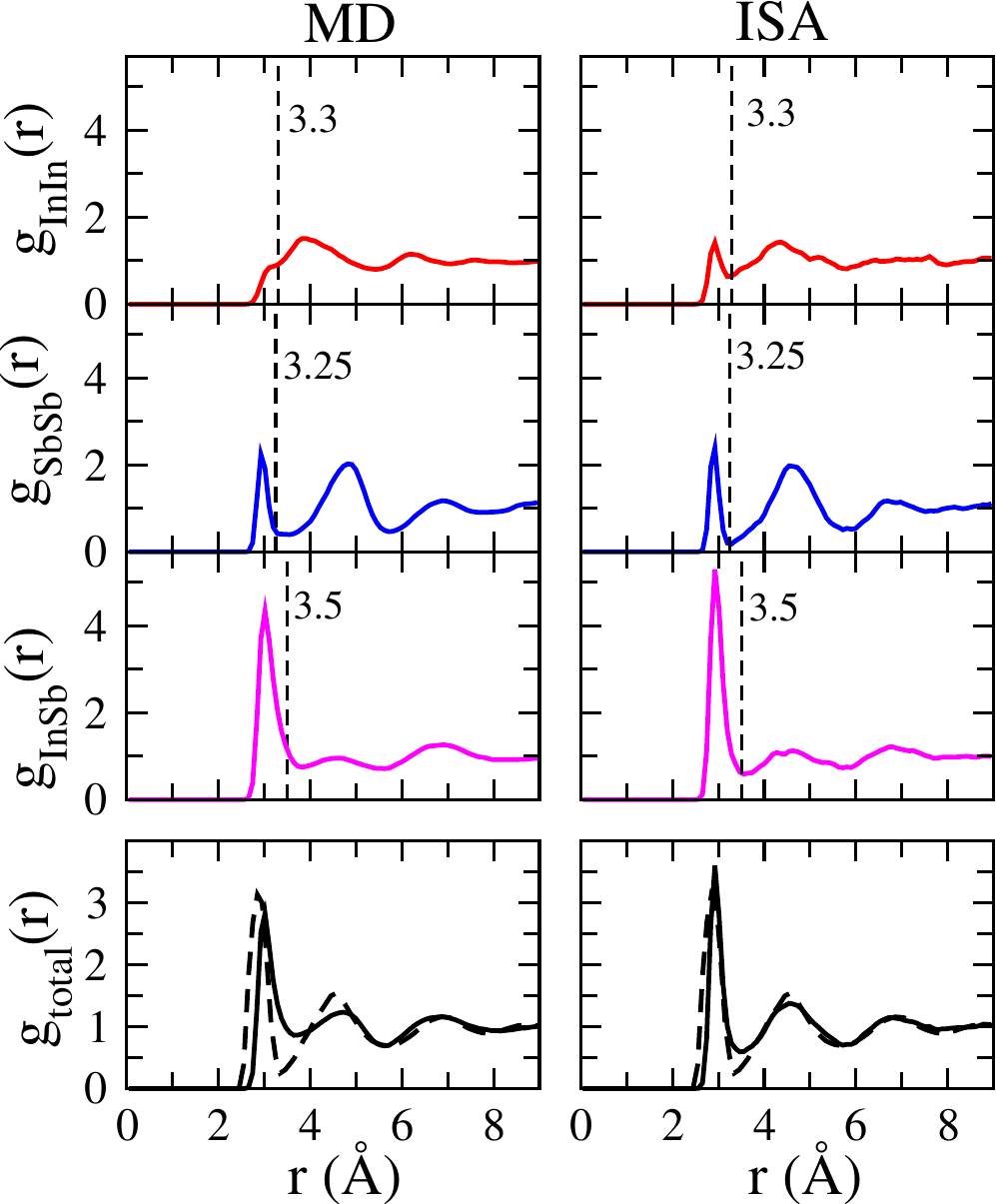}
\caption{\label{fig3:MDISA1} Comparison of the partial and total PCFs of the models of a-InSb, as obtained from a long DFT-based MD quench (MD-2) and the canonical ISA simulation. In the bottom graphs the experimental total $g(r)$ is given by dashed lines for comparison \protect\cite{Shevchik}. The vertical lines in the upper three panels are the corresponding cutoff radii, which are used in the following to define the coordination numbers.} 
\end{figure}

\begin{figure}
\vspace*{0.2cm}
\hspace*{0.15cm}
\includegraphics[width=7.5cm,clip]{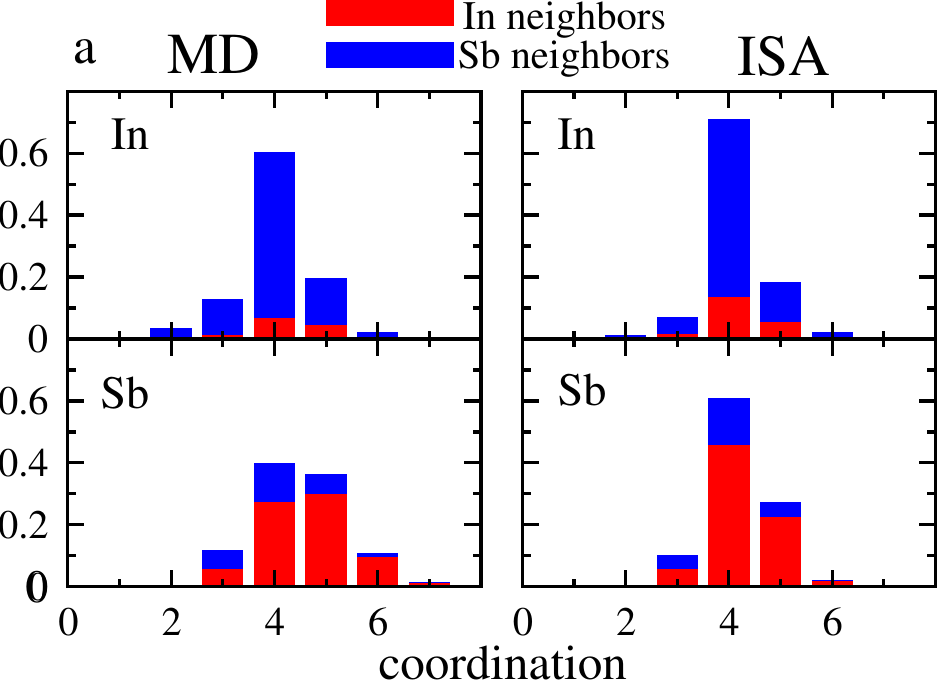}
\vspace*{0.05cm}
\hspace*{0.15cm}
\includegraphics[width=7.5cm,clip]{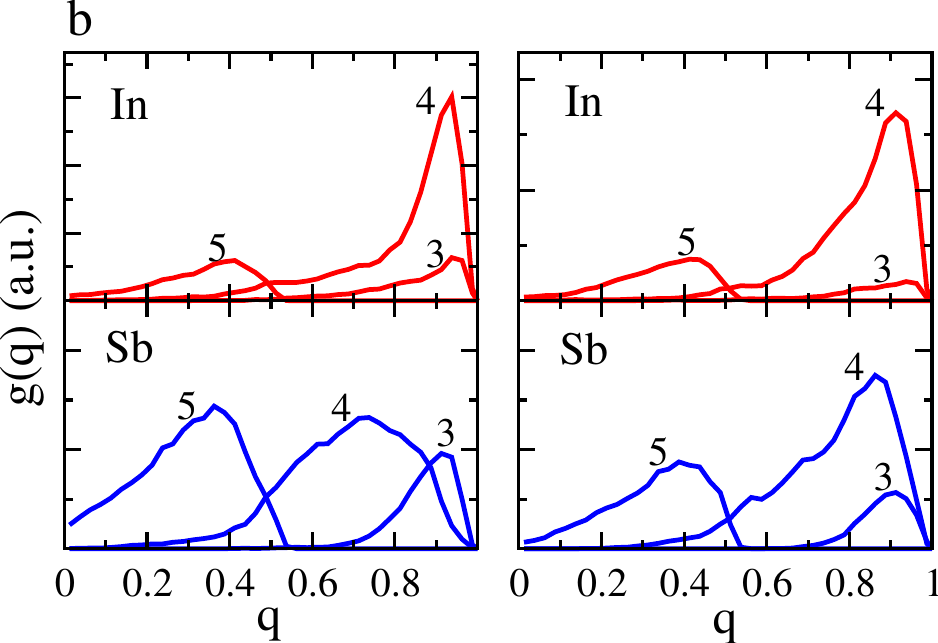}
\caption{\label{fig4:MDISA2} Comparison of the structural properties of a-InSb, as obtained from a long DFT-based MD quench and our modified canonical ISA method. In (a) the distribution of coordination numbers are shown, while (b) denotes the local order parameter of Eq.~\ref{qpar}.}
\end{figure}

In Figs.~\ref{fig3:MDISA1} and \ref{fig4:MDISA2} the structural properties of the energetically most favorable MD quench (MD-2) are compared with the model from the canonical ISA simulation. 
The partial and total PCFs, shown in Fig.~\ref{fig3:MDISA1}, were averaged over a 30~ps DFT-based MD trajectory at 300~K. That is to say that in the canonical ISA calculation, upon amorphization, the constraint has been removed. As shown in the bottom panels of Fig.~\ref{fig3:MDISA1}, the total PCF of the ISA model is in much closer agreement with the experimental $g(r)$ than the structure of the DFT-based MD quench. This immediately suggests that even after relieving the constraints, the ISA methods leads to models that are much less structured than those from rather long MD quenches and are generally much more reliable. 

In order to characterize the structures, the distribution of the coordination numbers and of the local order parameter $q$ are shown in Fig.~\ref{fig4:MDISA2}. The local order parameter $q$, introduced in Ref.~\onlinecite{Errington}, is an indicator of the tetrahedricity of the bonding geometry and is defined as:
\begin{eqnarray}
\label{qpar}
q_i = 1 - \frac{3}{8}
\sum_{j<k} \left(\frac{1}{3} + cos \theta_{ijk} \right)^2, 
\end{eqnarray}
where the sum runs over the pairs of atoms that are bonded to a central atom $i$ and are forming a bonding angle $\theta_{ijk}$. For a 3- or 4-fold coordinated tetrahedral environment $q$ = 1, while for 3-, 4-, 5- and 6-fold coordinated (defective) octahedral environments, $q$ evaluates to $\sim$0.87, $\sim$0.63, $\sim$0.33 and 0, respectively. To determine the neighboring atoms, the same cutoff distance values as shown in Fig.~\ref{fig3:MDISA1} have been used. 

As can be seen in Fig.~\ref{fig4:MDISA2}, the two structures we have considered are qualitatively differing from each other. 
The local environment of the 4-fold coordinated In atoms of the DFT-based MD quenches we generated is more and more tetrahedral the lower the eventual energy. The structure of the Sb atoms, however, is mainly defective octahedral, which leads to a too shallow first minimum in the total PCF. The structure of the canonical ISA simulation, however, is mostly 4-fold coordinated tetrahedral, similar to our previously published DFT-based MD quench \cite{Los2} and in overall very good agreement with the experimental PCFs.  

\subsection{Isobaric ISA Simulation}

Using the isobaric ISA method, the volume of the simulation cell is constantly adapted during the optimization in order to realize a predetermined target pressure. For the sake of simplicity have have confined ourselves to isotropic volume fluctuations \cite{NPTMD}, which permits to compute the particle density in cases where it is \textit{a priori} unknown, though the extension to also predict the cell shape similar to the Parrinello-Rahman scheme is straightforward \cite{PR}. 
However, contrary to the canonical ISA simulation, in the following an empirical van der Waals correction has been employed to better reproduce the experimental density \cite{Grimme}. 
At first, we equilibrated the liquid at 1000~K and constant ambient pressure using the isobaric hybrid MC technique without any constraints. At this temperature we found an an equilibrium density of 6.06~g/cm$^3$, which is 5 \% smaller than the associated experimental density \cite{Glazov,Chen}, but larger than the equilibrium density of the amorphous at 300~K. In fact, the average atomic coordination number in the liquid phase is also higher than in the amorphous, which is a common property of tetrahedral solids.

\begin{figure}
\vspace*{0.0cm}
\hspace*{0.15cm}
\includegraphics[width=7.5cm,clip]{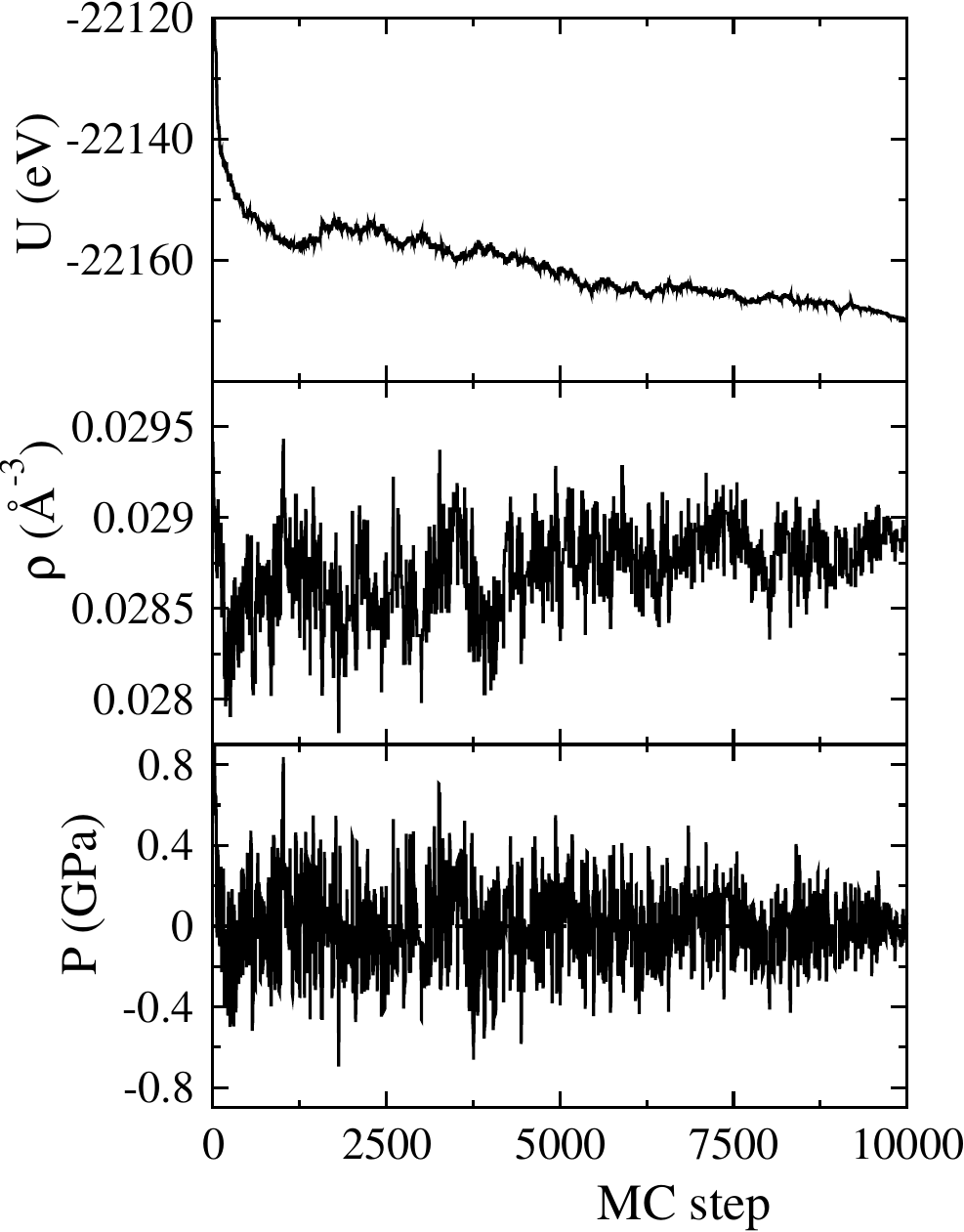}
\caption{\label{fig5:URhoP_ISA_VF} Time evolution of the potential energy $U$, the particle density $\rho$ and the pressure $P$ during an isobaric ISA simulation of InSb including isotropic volume fluctuations.}
\end{figure}

\begin{figure}
\vspace*{0.0cm}
\hspace*{0.15cm}
\includegraphics[width=7.5cm,clip]{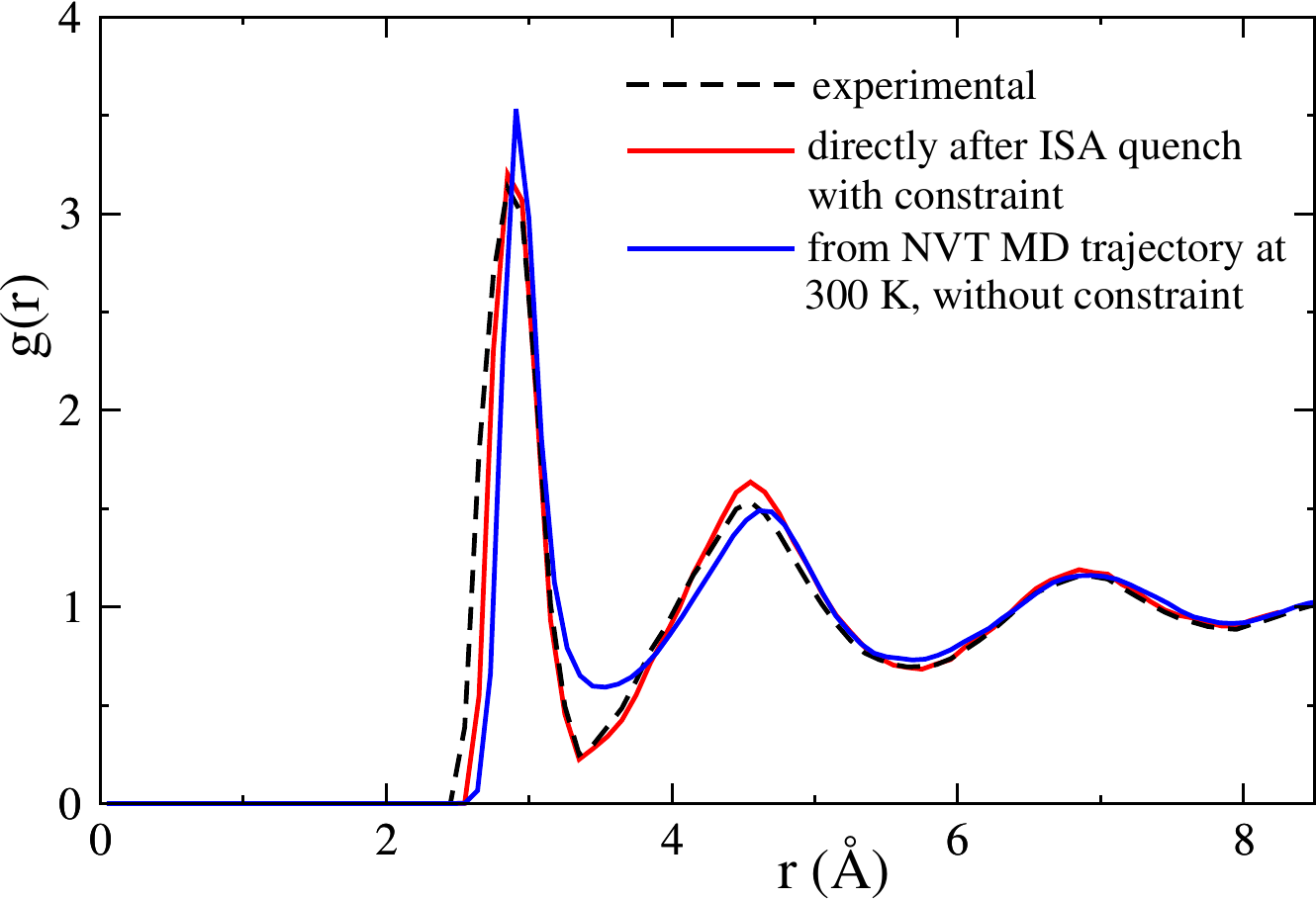}
\caption{\label{fig6:gr_ISA_VF} Total PCF of a-InSb, as obtained from a isobaric ISA simulation including van der Waals interactions, with and without the constraint. For the sake of comparison the experimental $g(r)$ from Ref.~\onlinecite{Shevchik} is shown.}
\end{figure}

\begin{figure}
\vspace*{0.1cm}
\hspace*{-0.05cm}
\includegraphics[width=7.7cm,clip]{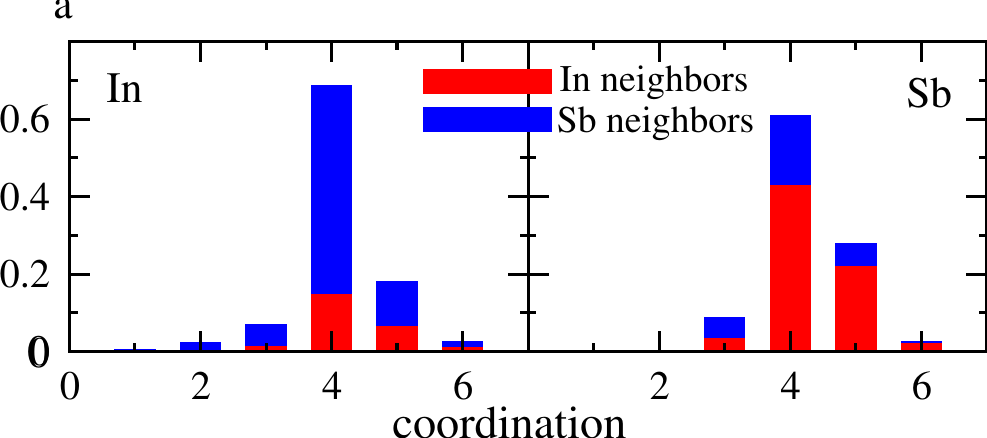}

\vspace*{0.1cm}
\hspace*{0.45cm}
\includegraphics[width=8.17cm,clip]{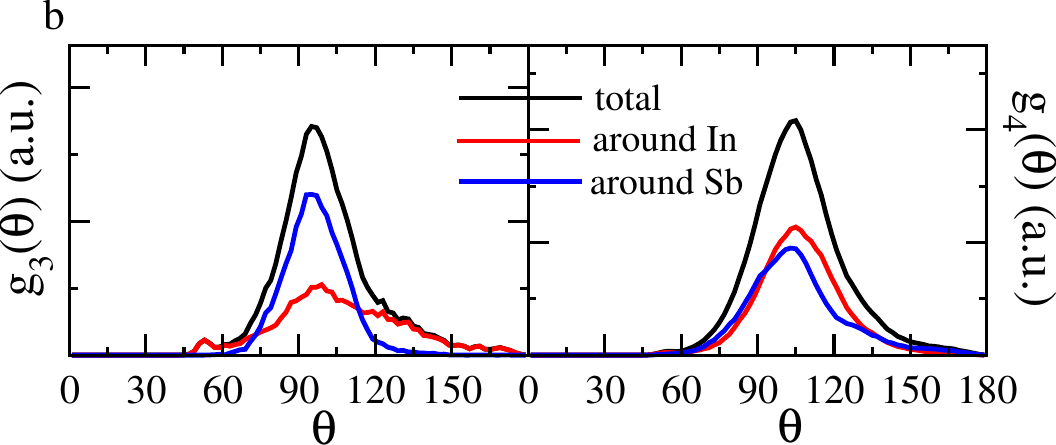}

\vspace*{0.0cm}
\hspace*{0.15cm}
\includegraphics[width=7.5cm,clip]{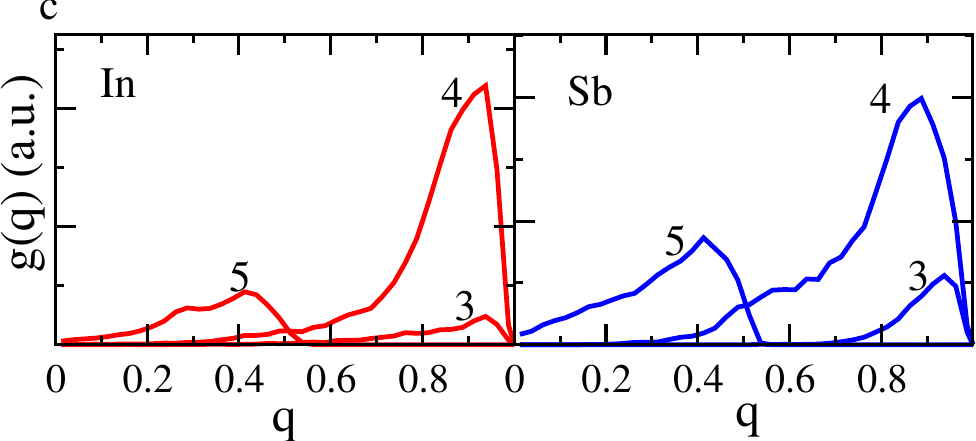}
\caption{\label{fig7:STR_ISA_VF} Structural properties of a-InSb generated by an isobaric ISA simulation. In (a) the distribution of coordination numbers are shown, while (b) exhibits the bond angle distribution and (c) the local order parameter of Eq.~\ref{qpar}.}
\end{figure}

The time evolution of the potential energy $U$, the particle density $\rho$ and the pressure $P$ during the isobaric ISA optimization is shown in Fig.~\ref{fig5:URhoP_ISA_VF}. 
The final density of the quenched amorphous is 5.67~g/cm$^3$, which is only 2 \% smaller than the experimental density \cite{Shevchik}. 
However, since this is only slightly larger than the previously estimated theoretical equilibrium density of 5.61~g/cm$^3$ \cite{Los2}, no appreciable changes due to van der Waals interactions are to be expected. In fact, the agreement of the experimental PCF and the computed $g(r)$ including the constraint, as shown in Fig.~\ref{fig6:gr_ISA_VF}, is equally excellent as in the case of the canonical ISA calculation. 
Likewise, even after relieving the constraint and equilibrating the amorphous at 300~K, the theoretical PCFs of the isobaric and canonical ISA simulations are in excellent agreement with each other, although the deviation from the experimental PCF is slightly larger than with the constraint. All other structural properties, shown in Fig.~\ref{fig7:STR_ISA_VF}, are also very similar to those obtained before using the canonical ISA method. 
The corresponding pair coordination numbers are given in Table~\ref{table_coor}.
\begin{table}
\begin{tabular}{ccccc}
\hline
 &   & with In     & with Sb     & total \\
\hline
&In  & 1.05 & 3.05 & 4.10 \\
&Sb  & 3.07 & 1.21 & 4.28 \\
\hline
\end{tabular}
\caption{Average pair coordination numbers of a-InSb computed from the partial PCFs of Fig.~\ref{fig6:gr_ISA_VF} as generated by an isobaric ISA simulation.}
\label{table_coor}
\end{table}

The present ISA results indicate that the structure of a-InSb that is compatible with the experimental PCF is mostly tetrahedral for both In and Sb atoms, which is in agreement with our previous work based on a long DFT-based MD quench using a rather confined basis set \cite{Los2}. Instead, quenching from the melt at the DFT-PBE level of theory employing a more accurate basis set than previously, but neglecting van der Waals interactions, the local structure of the 4-fold coordinated Sb atoms is more octahedral-like, although in less good agreement with experiment. 

\section{Conclusions}

We conclude by noting that the latter outcome points to a a potential deficiency of the employed PBE XC functional when dealing with the close competition in energy between tetrahedral-like and octahedral-like configurations, which is probably also responsible for the tetrahedral-to-octahedral transition observed experimentally in a-InSb under moderate pressure \cite{shimomurka}. Other XC functionals, however, such as the Becke-Lee-Yang-Parr (BLYP) \cite{Becke, LYP} or so-called hybrid functionals, which include some fraction of exact Hartree-Fock exchange, are known to entail a stronger electron localization than the employed PBE XC functional \cite{PBE}. A stronger electron localization is expected to favor tetrahedra against defective octahedra, as was indeed observed in simulations of liquid GeSe \cite{massobrio2}. It would therefore be interesting to assess the influence of the XC functional, including hybrids, on the structure of amorphous solids. 
This is now made possible by the superior efficiency of the ISA method, which allows to routinely determine the atomic structure of rather complex disordered systems at a higher level of theory, than previously thought feasible. 


\begin{acknowledgements}
We would like to thank the IDEE project of the Carl-Zeiss Foundation and the Graduate School of Excellence MAINZ for financial support.
\end{acknowledgements}

\section{Appendix: details on constraint term for the RPCF}
In the present ISA simulations the constraint on the RPCF, denoted by $\tilde{U}_G$ of Eq.~\ref{Ut_RPCF}, 
is calculated as:
\begin{eqnarray}
\label{Grn}
\hspace*{-0.6cm} 
G(r_n) = \frac{2}{ r_n \Delta r N } \sum_{<i,j>}
\int_{r_n - \frac{1}{2} \Delta r }^{r_n + \frac{1}{2} \Delta r } P_{ij} (r) dr
- 4 \pi r_n \rho
\end{eqnarray}
where $\Delta r = r_{n+1} - r_n$ and the sum over $<i,j>$ runs over all pairs of atoms, while $P_{ij} $ is a Gaussian-like polynomial of width $ w$ given by:
\begin{eqnarray}
\label{Pij}
P_{ij} (r) = \frac{15}{16w}
\left( 1 - \left( \frac{r-r_{ij}}{w} \right)^2 \right)^2, 
\end{eqnarray}
where $r$ is defined on the open interval $r \in (r_{ij} - w,r_{ij} + w)$, whereas $P_{ij} (r) = 0$ outside this interval. Here, $r_{ij}$ is the interatomic distance between the atoms $i$ and $j$. By construction, $\int_{-\infty}^{\infty} P_{ij} (r) dr = 1$ and $P_{ij} = dP_{ij}/dr = 0$ at $r = r_{ij} \pm w$, so that $P_{ij}$ is continuous up to the first derivative. Changing the variable to $x = (r-r_{ij})/w$ for a given pair pair of atoms $ij$, we rewrite: 
\begin{eqnarray}
\label{IntPij_1}
\int_{r_n - \frac{1}{2} \Delta r}^{r_n + \frac{1}{2} \Delta r} P_{ij} (r) dr &=&
\frac{15}{16} \int^{x_{max}}_{x_{min}} \left( 1 - x^2 \right)^2 dx \\ \nonumber 
&\equiv& Q_n(r_{ij}), 
\end{eqnarray}
where $x_{min} = max[ ( r_n - \frac{1}{2} \Delta r - r_{ij})/w, - 1 ]$, $x_{max} = min[ ( r_n - \frac{1}{2} \Delta r -r_{ij})/w, 1 ]$ and the analytic function $Q_n (r_{ij})$ a polynomial in $r_{ij}$.

The force contributions $ d\tilde{U}_G/dr_{i,\alpha}$ with $\alpha=x,y,z$ can now be computed as:
\begin{eqnarray}
\label{FUG}
\frac{d \tilde{U}_G}{dr_{i,\alpha}} &=&
2 w_G \sum_n ( G(r_n) - G^{exp} (r_n) )
\frac{d G(r_n)}{dr_{i,\alpha}}, 
\end{eqnarray}
where
\begin{eqnarray}
\label{dUGdri}
\frac{d G(r_n)}{dr_{i,\alpha}} = \frac{2}{ r_n \Delta r N } \sum_{<i,j>}
\frac{d r_{ij}}{dr_{i,\alpha}} \frac{d Q_n }{dr_{ij}}.
\end{eqnarray}
Applying a uniform scaling factor $s_r$ to scale the variables $r_n$, $\Delta r$ and $w$, results in $r'_n = s_r r_n$, $\Delta r'_n = s_r \Delta r_n$ and $w' = s_r w$, respectively. According to Eq.~\ref{Grn}, the scaled RPCF $G_s (r'_n)$ at $r'_n$ is equal to $G (r_n)/s_r^2$, where $G(r_n)$ is the RPCF before the scaling. Hence, $s_r^2 G_s (s_r r_n)$ is independent of the scaling factor $s_r$.

The optimal weight factor $w_g$ or $w_G$ depends on the number of grid points $N_{nr}$ in Eq.~\ref{Ut_PCF} or \ref{Ut_RPCF}, respectively, as well as on the choice of the smoothening parameter $w$ of Eq.~\ref{Pij}.
Varying the grid density, the ideal $w_G$ should scale as $1/N_{nr}$. A too small value for $w$ corresponds to an insufficient Gaussian smoothening and will give rise to spurious force contributions $d \tilde{U}_G/dr_{i,\alpha}$.

\end{document}